\documentstyle[11pt]{article}

\newcommand\tf{t_{\!f}}
\newcommand\tp{t_p}

\begin{document}

\title{{\bf Predicting Future Duration from Present Age:}\\
{\bf A Critical Assessment}\footnote{%
This work was supported in part by the U.S.~Office of Naval Research 
(Grant No.~N00014-93-1-0116).}}

\author{Carlton M.~Caves
\\
\\
Center for Advanced Studies\\
Department of Physics and Astronomy\\
University of New Mexico\\
Albuquerque, New Mexico \,87131-1156, USA\\
\\
E-mail: caves@tangelo.phys.unm.edu\\
Telephone: (505)\,277-8674
}

\date{21 January 2000}

\maketitle

\begin{abstract}
Using a temporal version of the Copernican principle, Gott has proposed a 
statistical predictor of future longevity based on present age 
[J.~R. Gott III, Nature {\bf 363}, 315 (1993)] and applied the predictor
to a variety of examples, including the longevity of the human species.  
Though Gott's proposal contains a grain of truth, it does not have the 
universal predictive power that he attributes to it.
\end{abstract}

\vglue 24pt

Returning from a five-week residence at the Isaac Newton Institute this
past summer, I found on my desk the July~21 issue of {\sl The New Yorker}, 
containing a provocative story by the well known science writer Timothy 
Ferris \cite{Ferris1999a}.  The story, entitled ``How to Predict Everything,'' 
describes how J.~Richard Gott, a Princeton astrophysicist, makes universal 
probabilistic predictions for a phenomenon's future duration based on knowing 
how long the phenomenon has lasted.  The justification for Gott's rule is 
said to be a {\it temporal\/} version of the Copernican principle: when you 
observe a phenomenon in progress, your observation does not occur at a 
special time.

Here is Gott's account, as related to Ferris, of how he conceived his rule
while contemplating the Berlin Wall.

\begin{quotation}
\small
Standing at the Wall in 1969, I made the following argument, using the 
Copernican principle. I said, Well, there's nothing special about the 
timing of my visit.  I'm just travelling---you know, Europe on five dollars 
a day---and I'm observing the Wall because it happens to be here.  My visit 
is random in time.  So if I divide the Wall's total history, from the 
beginning to the end, into four quarters, and I'm located randomly somewhere 
in there, there's a fifty-per-cent chance that I'm in the middle two 
quarters---that means, not in the first quarter and not in the fourth 
quarter. 

Let's suppose that I'm at the beginning of that middle fifty per cent.  In
that case, one quarter of the Wall's ultimate history has passed and there 
are three quarters left in the future.  In that case, the future's three 
times as long as the past.  On the other hand, if I'm at the other end, 
then three quarters have happened already, and there's one quarter left in 
the future.  In that case, the future is one-third as long as the past. 

\dots\ 

(The Wall was) eight years (old in 1969).  So I said to a friend, ``There's 
a fifty-per-cent chance that the Wall's future duration will be between 
(two and) two-thirds of a year and twenty-four years.''  Twenty years later, 
in 1989, the Wall came down, within those two limits that I had predicted.
I thought, Well, you know, maybe I should write this up.
\end{quotation}

\noindent
Ferris goes on to recount how Gott applies his method to the longevity of 
the human species.  

\begin{quotation}
\small
The question that Gott has been asking lately is how long the human species
is going to last.  Since scientists generally make predictions at the 
ninety-five-per-cent confidence level, Gott begins with the assumption that 
you and I, having no reason to think we've been born in a special time, are 
probably living during the middle ninety-five per cent of the ultimate duration 
of our species.  In other words, we're probably living neither during the first 
two and a half per cent nor during the last two and a half per cent of all 
the time that human beings will have existed.  

``{\it Homo sapiens\/} has been around for two hundred thousand years,''
Gott said \dots\,.  ``That's how long our past is.  Two and half per cent 
is equal to one-fortieth, so the future is probably at least one-thirty-ninth 
as long as the past but not more than thirty-nine times the past.  If we 
divide two hundred thousand years by thirty-nine, we get about fifty-one 
hundred years.  If we multiply it by thirty-nine, we get 7.8 million 
years.  So if our location in human history is not special, there's a 
ninety-five-per-cent chance we're in the middle ninety-five per cent of 
it. Therefore the human future is probably going to last longer than 
fifty-one hundred years but less than 7.8 million years.

``Now, those numbers are interesting, because they give us a total longevity 
that's comparable to that of other species.''
\end{quotation} 

These glib predictions astonished me, not because Gott concludes from them 
that {\it homo sapiens\/} is unlikely to last longer than other 
species---that is a legitimate subject for inquiry and debate---but
because they are put forward as a universal rule, applicable no matter 
what other information one has about the phenomenon in question.  In 
making statistical predictions of future longevity, Gott dismisses 
the entire process of assembling and organizing information about a
phenomenon, evaluating that information critically, and if possible, 
formulating laws that describe the phenomenon.  Put succinctly, he 
rejects as irrelevant the process of rational, scientific inquiry, 
replacing it with a single, universal statistical rule.  That has to be 
wrong.  

I decided it was important to find the flaws in Gott's reasoning: flawed
thinking is an inevitable, even necessary part of the scientific enterprise,
but when it makes its way into {\sl The New Yorker}, the time has come to 
find the flaws and draw attention to them.  I began by requesting from 
the UNM Library a copy of the {\sl Nature\/} article \cite{Gott1993a} where 
Gott proposes his rule and applies it to the above examples and others.  A 
citation search turned up two other pieces in which Gott adds to the content 
of his {\sl Nature\/} article: a Letter to {\sl Nature\/} \cite{Gott1994a} 
responding to letters criticizing the original article and a chapter in 
the proceedings of an Astronomical Society of the Pacific (ASP) Symposium 
\cite{Gott1996a}.  The present paper analyzes what I found in Gott's papers 
and reports my conclusions.

\bigbreak

\leftline{\bf Gott's delta-t argument}

\nobreak
\smallskip

Gott justifies his probabilistic predictions by making what he calls the 
delta-$t$ argument \cite{Gott1993a}.  Suppose there is a phenomenon that has 
a beginning, or birth, at time $t_0$ and an end, or death, at time $t_0+T$, 
$T$ being the duration of the phenomenon.  You observe the event at a time 
$t$ between the beginning and the end, corresponding to a present age, 
$\tp=t-t_0$, and a future duration, $\tf=T-\tp$.  If there is nothing special 
about the observation time---this is the content of the temporal 
Copernican principle---Gott reasons that $\tp=t-t_0$ is a random variable 
uniformly distributed between 0 and $T$.  This means that $\tp$ lies 
between $aT$ and $bT$, $0\le a\le b \le1$, with probability $b-a\equiv f$; 
in symbols, we write
\begin{equation}
P(aT<\tp<bT)=b-a=f\;.
\label{eq:Copernicus}
\end{equation}

Gott's next step is to infer from Eq.~(\ref{eq:Copernicus}) that the duration 
$T$ lies between the corresponding bounds, $\tp/b$ and $\tp/a$, with the 
same probability $f$.  Translated to future duration, this says that $\tf$ 
lies between $(b^{-1}-1)\tp$ and $(a^{-1}-1)\tp$ with probability $f$, i.e., 
\begin{equation}
G\Biggl({1-b\over b}\tp<\tf<{1-a\over a}\tp\Biggr)=b-a=f\;.
\label{eq:Grule1}
\end{equation}
All of Gott's predictions flow from this probability rule.  I use the letter 
$G$ to distinguish probabilities based on this rule.

Gott phrases his predictions in terms of particular $f\times100\%$ confidence 
levels, which he obtains by letting $a$ and $b$ be equidistant from 0 and 1, 
i.e., $a=1-b$.  The resulting choices, $a={1\over2}(1-f)$ and 
$b={1\over2}(1+f)$, lead to Gott's confidence-level prediction:
\begin{equation}
\matrix{
\displaystyle{{1-f\over1+f}\tp<\tf<{1+f\over1-f}\tp}\cr
\vphantom{\biggl(}\mbox{($f\times100\%$ confidence level)}
}\;.
\label{eq:Grule2}
\end{equation} 
For example, in his encounter with the then $(\tp\,\mathord{=}\,)$8-year-old
Berlin Wall, Gott used $f=1/2$, with $a=1/4$ and $b=3/4$, which led him to 
predict with 50\% confidence that the total duration of the Wall would lie 
between $4\tp/3=10{2\over3}\,$yr and $4\tp=32\,$yr or, equivalently, that 
the future duration would lie between $\tp/3=2{2\over3}\,$yr and $3\tp=24\,$yr.  
In most of his work, Gott uses a 95\% confidence level, corresponding to 
$f=0.95$.   

Another form of Gott's rule arises from letting $b=1$ and $a=(1+Y)^{-1}$.  
Inserting these choices into Eq.~(\ref{eq:Grule1}), one finds that 
$\tf<Y\tp$ with probability $Y/(1+Y)$; equivalently, the probability that 
the future duration is not less than $Y\tp$ is $(1+Y)^{-1}$, 
i.e., 
\begin{equation}
G(\tf\ge Y\tp)={1\over1+Y}\;.
\label{eq:Grule3}
\end{equation}
In his {\sl Nature\/} article, Gott derives Eq.~(\ref{eq:Grule3}) independently 
of the delta-$t$ argument by assuming that the phenomenon of interest is 
an exponential decay [see Gott's Eq.~(6) and preceding discussion].  There 
being no hint in the delta-$t$ argument that Gott restricts his 
method to exponential decays, this derivation must be intended as an 
{\it example\/} of his method.  I defer discussion of this derivation, since 
its status can be appreciated only after exposing and correcting the flaws 
in Gott's reasoning.

The delta-$t$ argument implies that Gott's rule provides a universal method
for predicting the future duration of any phenomenon, the only assumption 
being that the observation time is not special.  Moreover, it is clear from 
the variety of phenomena to which Gott applies his rule---durations of the 
Berlin Wall, Stonehenge, and the Soviet Union, the publication lifetime of 
{\sl Nature}, longevity of the human species, and in his ASP contribution and
in his conversations with Ferris, running times of plays in New York---that 
he places no restrictions on the applicability of his rule.

It is not hard to find an error in the delta-$t$ argument: {\it the step 
from Eq.~(\ref{eq:Copernicus}) to Gott's rule~(\ref{eq:Grule1}) has no 
justification in probability theory.}  This error that has been pointed 
out by Buch, in a Letter to {\sl Nature\/} criticizing Gott's method 
\cite{Buch1994a}.  The total duration $T$ (or the future duration~$\tf$) 
is unknown and thus must be treated as a random variable described by a 
prior probability distribution.  This prior distribution expresses whatever 
information one possesses that can be used to make probabilistic statements 
about the phenomenon's duration.  After collecting the data that the 
phenomenon's present age is $\tp$, the only procedure authorized by 
probability theory is to update the prior distribution to a new, posterior 
distribution for $T$ (or $\tf$), which reflects both the prior information 
and the present age.  The formal procedure for this updating is called 
Bayes's theorem~\cite{Bernardo1994a}.  

The error just identified is sufficient to invalidate the delta-$t$ argument.  
To correct it requires an analysis that uses Bayes's theorem to update 
probabilities.  Indeed, Gott has endorsed \cite{Gott1994a,Gott1996a} a
Bayesian analysis suggested by Buch \cite{Buch1994a}; this Bayesian analysis, 
said to be based on the temporal Copernican principle, leads to Gott's 
rule, provided one uses a particular prior distribution, $dT/T$, called the 
Jeffreys prior \cite{Jeffreys1939a}.  The reader should be aware, however, 
that the Bayesian analysis suggested by Buch and endorsed by Gott is also 
flawed.  In considering the Buch-Gott Bayesian analysis below, we will 
uncover this flaw, thus revealing a second error in the delta-$t$ argument, 
just as serious as the first, but more insidious because it is more subtle: 
{\it Equation~(\ref{eq:Copernicus}) is an incorrect mathematical formulation 
of the temporal Copernican principle.}  The pay-off for identifying this 
second flaw is that it clarifies the meaning and status of the temporal 
Copernican principle.  In developing a proper Bayesian analysis based on 
the temporal Copernican principle, we will discover that Gott's rule is 
a universal consequence of the Copernican principle, {\it in the situation 
where one knows the phenomenon to be in progress, but does not know its 
present age.}  Not knowing the present age, one cannot make Gott's 
predictions of future duration.

Before turning to the Bayesian analysis, however, I introduce a few examples 
that show that Gott's rule cannot be a universal predictor and also serve 
to put some flesh on the dry bones of the subsequent Bayesian analysis.

\bigbreak

\leftline{\bf Examples of using Gott's rule}

\nobreak
\smallskip

I advise my students to test the solution to a homework problem by considering 
special cases where the solution is already known.  This common-sense 
technique, a good rule in scientific thinking and in everyday life, 
provides compelling evidence that Gott's predictions cannot have the 
universal validity that he attributes to them.

\begin{itemize}

\item
{\it Exponential decay.} Consider an atom that is excited to a metastable 
energy level at some unknown time and then decays exponentially to the 
ground state with a decay constant $\tau^{-1}=(20\,{\rm min})^{-1}$.  
You come along at time $t$ and are told that the atom is in the metastable 
level, having been excited a time $\tp=15\,$min ago.  According to Gott, 
you can predict with 95\%\ confidence that the decay will occur between 
$\tf=\tp/39=23.1\,$s and $\tf=39\tp=9.75\,$hr into the future; more 
telling is that Eq.~(\ref{eq:Grule3}) predicts that $\tf\ge 4\tp=60.0\,$min 
with probability 1/5.  These predictions contradict the defining property 
of an exponential decay: being informed that the atom is still in the 
excited state at time $t$ simply resets the clock so that your expectations 
for its future decay are the same as though it had been initially excited 
at time $t$.  Specifically, you predict that it will survive a further time 
$\tf$ without decaying with probability $e^{-\tf/\tau}$, corresponding to 
95\% confidence of decay between $\tf=0$ and $\tf=\tau\ln20=3.00\tau=59.9\,$min.
Though the numerical discrepancies between Gott's predictions and the 
predictions of an exponential decay are important, they are only a symptom 
of the real problem: Gott's rule, by including present age in the prediction 
of future duration, is inconsistent with the very notion of an exponential 
decay.

\end{itemize}

Buch \cite{Buch1994a} has pointed out that Gott's rule is inconsistent with 
the properties of an exponential decay.  In his reply to Buch \cite{Gott1994a}
and in his ASP contribution \cite{Gott1996a}, Gott admits that his method
doesn't apply to an exponential decay whose decay constant is known.  
Instead, he says that it applies to an exponential decay whose decay constant 
is unknown and distributed according to the Jeffreys prior $d\tau/\tau$; this 
leads to the Jeffreys prior $dT/T$ for total duration $T$ and is not an 
exponential decay at all.  Gott \cite{Gott1994a} also reasserts his 
exponential-decay derivation of Eq.~(\ref{eq:Grule3}), to be discussed below.  
All this leaves one thoroughly confused---does Gott regard his rule as 
universal or not?---but his subsequent conversations with Ferris 
\cite{Ferris1999a} make clear that he does not acknowledge any restrictions 
on the use of his rule.

\begin{itemize}

\item
{\it Longevity of an individual.}  Suppose you are going to a meeting 
of your book club, to be held at a member's house that you've never 
been to before.  You find the right street, but having forgotten the 
street address, you choose between two houses where there is evident 
activity.  Knocking at one, you are told that the activity within 
is a birthday party, not a book-club meeting.  Your friendly enquiry about 
the age of the celebrant elicits the reply that she is celebrating her 
$(\tp\,\mathord{=}\,)$50th birthday.  According to Gott, you can predict with 
95\% confidence that the woman will survive between $\tp/39=1.28\,$years and 
$39\tp=1,950\,$years into the future.  Since the wide range encompasses 
reasonable expectations regarding the woman's survival, it might not seem 
so bad, till one realizes that Eq.~(\ref{eq:Grule3}) predicts that with 
probability 1/2 the woman will survive beyond 100 years old and with 
probability 1/3 beyond 150.  Few of us would want to bet on the woman's 
survival using Gott's rule.

\end{itemize}

One might object at this point that Gott probably didn't intend his rule to 
apply to an individual's longevity, but in his ASP contribution 
\cite{Gott1996a}, Gott applies the rule to himself: ``At the time my 
({\sl Nature\/}) paper was published on May~27, 1993, I was 46.3 years old, 
so the 95\% delta-$t$ argument predicted that my future longevity would be 
at least 1.2 years but less than 1,806 years.  I have survived past the 
lower limit already and so if I don't make it past the upper limit, then 
that prediction will indeed prove correct for me!''

\begin{itemize}

\item
{\it Deterministic phenomena.} The best testing ground for ideas comes 
from extreme cases, and here the most extreme case is a deterministic 
phenomenon.  Putting the example in a dramatic context, suppose you are 
captured by terrorists, who confine you to a small room.  You are told 
that at some time in the next 24 hours, a timer will be set and that after 
it has ticked for 30 minutes, poison gas will fill the room, killing you.  
You are then drugged and wake up to find the timer ticking and reading 
20~minutes since being set.  According to Gott, you can predict with 
95\%\ confidence that the time to release of the gas lies between 
$\tf=20\,{\rm min}/39=30.8\,$s and $\tf=39\times20\,{\rm min}=13.0\,$hr; 
even worse, Eq.~(\ref{eq:Grule3}) predicts that with probability 2/3 the 
time to release is $10\,$min or more.  These reassuring predictions provide 
scant comfort, since you {\it know\/} you have $10\,$min to live. 

\end{itemize}

These examples demonstrate that Gott's rule cannot be a universal 
method for predicting future durations.  If the rule has any validity, it 
must involve other information than the present age of a phenomenon.  As 
E.~T. Jaynes taught us \cite{Jaynes1986a}, when probabilistic predictions 
violate one's intuition, the proper response is neither to accept the 
nonintuitive predictions without question nor to dismiss them out of hand, 
but rather to identify the information underlying the prediction.  You will 
either find the information inapplicable to the situation at hand, thereby 
confirming your intuition and allowing you to discard the predictions, or 
you will sharpen your intuition.  

The objective of this paper is to identify the prior information that underlies
Gott's rule.  The tool is Bayesian analysis.  We will discover that the 
temporal Copernican principle contains a grain of truth, but that grain 
of truth does not include Gott's predictions of future duration.

\bigbreak

\leftline{\bf A flawed, but instructive Bayesian analysis}

\nobreak
\smallskip

Return to the general situation introduced above, that of a phenomenon 
with a birth time $t_0$ and a duration $T$.  You observe the phenomenon at 
time $t$.  It is often useful to replace one or both quantities, $t_0$ and 
$T$, by the present age, $\tp=t-t_0$, and the future duration, $\tf=T-\tp$.  
In developing the Bayesian analysis, I first formulate and analyze a
flawed approach, advanced by Buch \cite{Buch1994a} and endorsed by Gott 
\cite{Gott1994a}, which is modeled on Gott's delta-$t$ argument.  For this 
purpose, it is most convenient to use $\tp$ and $T$ as the primary variables.  
The reason for going through this flawed analysis is that it turns up 
the second error in Gott's delta-$t$ argument. 

Your prior information about the phenomenon is expressed in a prior 
probability $p(\tp,T)\,d\tp\,dT$, the joint probability that the phenomenon
has lasted a time between $\tp$ and $\tp+d\tp$ at the time of observation 
and that the phenomenon will last a total time between $T$ and $T+dT$.   
The joint probability density can be written as $p(\tp,T)=p(\tp|T)w(T)$, 
where $p(\tp|T)$ is the {\it conditional\/} probability density for the 
present age, given a total duration $T$, and $w(T)$ is your prior probability 
density for the total duration.  Throughout I use upper-case letters for 
probabilities and lower-case letters for probability densities.

Before going further, it is useful to introduce two quantities related to
$w(T)$: $\lambda(T)$ is the death rate---i.e., $\lambda(T)\,dT$ is the 
probability that the phenomenon, having lasted a time $T$, ends in the 
next $dT$---and $Q(T)$ is the survival probability---the probability that 
the phenomenon lasts at least a time $T$.  These quantities are related by
\begin{equation}
w(T)=-{dQ\over dT}=Q(T)\lambda(T)
\end{equation}
or, equivalently, by
\begin{equation}
Q(T)=\int_T^\infty dT'\,w(T')=\exp\!\left(-\int_0^T dT'\,\lambda(T')\right)
\;.
\end{equation}
An exponential decay is characterized by a constant death rate, 
$\lambda(T)=\lambda_0$, in which case $Q(T)=e^{-\lambda_0T}$ and 
$w(T)=\lambda_0e^{-\lambda_0T}$.

Gott's formulation of the temporal Copernican principle is the following:
if there is nothing special about the observation time, the present age is
a random variable uniformly distributed between 0 and $T$, i.e., 
\begin{equation}
p(\tp|T)\,d\tp=d\tp/T\;,\quad 0\le\tp\le T.
\label{eq:uniformtp}
\end{equation}
This is the probability-density version of Eq.~(\ref{eq:Copernicus}).  We 
now use Bayes's theorem, 
\begin{equation}
P(X|Y)P(Y)=P(X,Y)=P(Y|X)P(X)\;,
\end{equation}
to find your posterior probability density for the total duration, given 
the present age:
\begin{equation}
p(T|\tp)={p(\tp|T)w(T)\over p(\tp)}=
\cases{
	0\;,&$T<\tp$,\cr
	w(T)/Tp(\tp)\;,&$T\ge \tp$.
      }
\label{eq:wrongpT}
\end{equation}
The unconditional probability density $p(\tp)$ for the present age, which 
is a normalization constant in this expression, is given by
\begin{equation}
p(\tp)=\int_{\tp}^\infty dT\,{w(T)\over T}\;.
\end{equation}

One can easily verify that Gott's rule, embodied in 
Eqs.~(\ref{eq:Grule1})--(\ref{eq:Grule3}), is equivalent to a posterior 
density 
\begin{equation}
p(T|\tp)=
\cases{
        0\;,&$T<\tp$,\cr
        \tp/T^2\;,&$T\ge \tp$.
      }  
\end{equation}
To get this posterior from the present analysis, one must assume the 
(unnormalizable) Jeffreys prior, 
\begin{equation}
w(T)={1\over T}\;. 
\end{equation}
Buch \cite{Buch1994a} concludes that Gott's rule is unreasonable because 
it corresponds to an unnormalizable prior density.  Gott \cite{Gott1994a} 
replies (correctly, I think) that there is nothing wrong with an 
unnormalizable prior, since the posterior density for $T$ can be normalized.  
He defends the Jeffreys prior as being the appropriate ``vague Bayesian 
prior'' to use in a situation where one initially knows nothing about the 
magnitude of the duration \cite{Gott1994a,Gott1996a}.  

Jaynes \cite{Jaynes1968a} has delineated the conditions for using the 
Jeffreys prior, showing that it should be used when one's prior information 
is unchanged by a rescaling of the total duration, $T'=\alpha T$ ($\alpha>0$).  
If one's prior information is unchanged by the rescaling, then the density 
for $T'$, $w'(T')=w(T)dT/dT'=w(T)/\alpha$, should have the same functional 
form as the original density, i.e., $w'(T')=w(T')$.  This gives 
$w(\alpha T)=w(T)/\alpha$, which implies that $w(T)\propto1/T$.  This might 
seem to be progress in identifying the information that underlies Gott's 
rule---use it when one has no prior information about time scales associated 
with the phenomenon---but it turns out not to be, because the present Bayesian 
analysis is wrong.  The reason for presenting it is not to consider its 
consequences, but to identify where it goes wrong.

\bigbreak

\leftline{\bf A straightforward Bayesian analysis}

\smallskip

That something is wrong is made apparent by a different analysis of the 
same situation, this time a straightforward Bayesian analysis that does 
not invoke the Copernican principle.  Your prior information about 
the total duration is expressed in the prior density $w(T)$.  You observe 
the phenomenon still to be in progress a time $\tp$ after its beginning.  
The conditional probability for this observation, given a total duration 
$T$, is 0 if $\tp>T$ and 1 if $\tp\le T$.  Thus Bayes's theorem implies,
with $O$ denoting the observation,
\begin{equation}
p(T|O)=
{P(O|T)w(T)\over P(O)}
=\cases{
	0\;,&$T<\tp$,\cr
	w(T)/Q(\tp)\;,&$T\ge\tp$.
       }
\label{eq:rightpT}
\end{equation}
Here the normalization constant is the survival probability, i.e., $P(O)=Q(\tp)$.

The posterior density~(\ref{eq:rightpT}) is so eminently reasonable that one
could have written it down without using the formal apparatus of Bayes's 
theorem.  It says that the effect of discovering the present age is to 
rule out durations shorter than the present age; your posterior expectations
for durations longer than the present age are the same as your prior 
expectations, with appropriate renormalization.  Notice that this inference
updates sensibly: subsequent observations that find the phenomenon still 
in progress simply exclude a wider interval of durations.  Yet putting this 
simple inference in the context of the Copernican principle apparently yields 
a different posterior density~(\ref{eq:wrongpT}) for the total duration.  
How can that be?  There's nothing wrong with the Bayesian inference in either
analysis, so the culprit must be Gott's formulation of the temporal Copernican 
principle.  Thus we arrive at the second error in the delta-$t$ argument: 
{\it the uniform density~(\ref{eq:uniformtp}) for $\tp$---and, by extension, 
Eq.~(\ref{eq:Copernicus})---is not the correct mathematical formulation 
of the temporal Copernican principle.}

Where the uniform density goes wrong is in {\it assuming\/} that your 
observation occurs while the phenomenon is in progress.  If your observation 
does not occur at a special time, then it is very likely that it occurs before 
the phenomenon begins or after it has ended.  Including these other 
possibilities leads to a proper Bayesian formulation of the temporal 
Copernican principle, which is consistent with the inference expressed 
in Eq.~(\ref{eq:rightpT}).

\bigbreak

\leftline{\bf A proper Bayesian analysis of the temporal Copernican principle}

\nobreak
\smallskip

In formulating a proper Bayesian analysis, it is convenient to choose the
birth time $t_0$ and the total duration $T$ as the primary variables.
Your prior knowledge about these two quantities is incorporated in two 
probability densities: (i)~$\gamma(t_0)$ gives the probability 
$\gamma(t_0)\,dt_0$ that the phenomenon begins between times $t_0$ and 
$t_0+dt_0$; (ii)~$p(T|t_0)$ gives the probability $p(T|t_0)\,dT$ that 
the phenomenon lasts a time between $T$ and $T+dT$, given that it began 
at time $t_0$.  The corresponding joint probability density is 
$p(t_0,T)=p(T|t_0)\gamma(t_0)$.

The temporal Copernican principle---that your observation does not take
place at a special time---is a time-translation symmetry that restricts 
the form of the prior densities \cite{Jaynes1968a}.  To say that your 
observation time is not special is to say that your prior information 
is unchanged if the entire phenomenon is displaced in time while your 
observation time remains fixed.  To be consistent with this translation 
symmetry, your prior probability density should be unchanged by such a 
time translation; i.e., $p(t_0,T)$ should be independent of the birth time 
$t_0$.  Thus the temporal Copernican principle can be captured precisely in 
the following two statements:

\begin{enumerate}

\item
The phenomenon is equally likely to begin at any time.  This means that
$\gamma(t_0)$ is a constant.  In order to work with normalizable 
probabilities, I replace the exact symmetry with the approximate one that 
$\gamma(t_0)$ has a constant value, $1/\Delta$, at all times within a 
very long time interval.   The duration $\Delta$ of this very long time
interval exceeds all other times relevant to the problem, particularly
typical durations.

\item
Probabilities for total duration are independent of birth time.  This 
means that the conditional probability density $p(T|t_0)$ does not depend 
on $t_0$ and can be written as $p(T|t_0)=w(T)$, where $w(T)$ is the 
probability density introduced above.

\end{enumerate}

Should you be dissatisfied with these restrictions on 
the prior probabilities, it means that you do not accept the temporal
Copernican principle as applying to your prior information.  Dissatisfaction
should not be surprising, for one would not expect the Copernican principle 
to apply to all situations.  The three examples introduced above illustrate 
considerations that arise in using the temporal Copernican principle.  In 
all three examples, it is easy to accept that ignorance of the birth time 
is described by the time-translation symmetry of the temporal Copernican 
principle: the atom can be excited at any time during an interval much longer 
than the decay time; for the woman at the birthday party, the situation could 
be phrased in terms of an individual whose birth could occur at any time over 
a period much longer than a typical human lifetime; the timer can be set at 
any time within a 24-hour period, a period somewhat longer than the 30 minutes 
that the timer ticks.  Moreover, in the cases of the atom and the poison gas, 
duration probabilities are independent of the birth time.  In contrast, in 
the case of the longevity of an individual, the prior conditional probability 
for the individual's lifetime would depend on the time of birth.  Your prior 
expectation for the longevity of an individual born, say in Britain, would 
depend on whether the individual was born in the second half of the 20th 
Century, at the beginning of the 19th Century, or 10,000~years ago, at the 
end of the last Ice Age.  

At time $t$ you make your observation.  In Gott's formulation the observation
yields the present age, but we now understand that getting the present age 
presupposes that your observation finds the phenomenon in progress.  The 
first result of the observation is simply to determine whether the phenomenon 
has not yet begun, is already over, or is in progress.  Only the last of these 
possibilities, denoted by $I$ for ``in progress,'' is of interest to us.  
The conditional probability to find the phenomenon in progress, given a birth 
time $t_0$ and a duration $T$, is 
\begin{equation}
P(I\,|t_0,T)=\cases{
	1\;,&$t_0\le t\le t_0+T$,\cr
	0\;,&otherwise.
		 }
\end{equation}
The unconditional probability to find the phenomenon in progress is given by
\begin{eqnarray}
P(I)&=&
\int_{-\infty}^\infty dt_0\int_0^\infty dT\,
P(I\,|t_0,T)\gamma(t_0)w(T)\nonumber\\
&=&\int_{-\infty}^t dt_0\,\gamma(t_0)
\underbrace{\int_{t-t_0}^\infty dT\,w(T)}_{\displaystyle{\mbox{}=Q(t-t_0)}}\\
&=&\int_0^\infty dt'\,\gamma(t-t')Q(t')\;.\nonumber
\end{eqnarray}
The assumption that $\gamma(t_0)$ is constant for all times of interest
means that $\gamma(t-t')=1/\Delta$ for all times $t'$ such that the survival
probability $Q(t')$ is significantly different from zero.  This allows us to 
put $P(I)$ in the form
\begin{equation}
P(I)=\overline T/\Delta\;,
\end{equation}
where
\begin{equation}
\overline T=\int_0^\infty dT\,Tw(T)=\int_0^\infty dT\,Q(T)
\end{equation}
is the mean total duration with respect to the prior density $w(T)$.  The 
present analysis assumes that $\overline T$ is finite, which requires, for 
large durations $T$, that $Q(T)$ go to zero faster than $1/T$ or, equivalently, 
that $w(T)$ go to zero faster than $1/T^2$.  For an exponential decay, 
$\overline T^{-1}=\lambda_0$ is the decay constant.  Notice that the 
probability to find the phenomenon in progress is very small.

Bayes's theorem gives the posterior probability density for $t_0$ and $T$, 
given that the phenomenon is occurring:
\begin{equation}
p(t_0,T|I)={P(I|t_0,T)\gamma(t_0)w(T)\over P(I)}
=\cases{\gamma(t_0)w(T)/P(I)\;,&$t_0\le t\le t_0+T$,\cr
	0\;,&$t_0>t$ or $t_0+T<t$.
       }
\end{equation}
If $t-t_0$ is large enough in this expression that $\gamma(t_0)$ does not have 
its constant value, then $T\ge t-t_0$ is so large that $w(T)$ is negligible.
Thus we can again replace $\gamma(t_0)$ by the constant value $1/\Delta$,
leaving
\begin{equation}
p(t_0,T|I)
=\cases{w(T)/\overline T\;,&$t_0\le t\le t_0+T$,\cr
	0\;,&$t_0>t$ or $t_0+T<t$.
       }
\label{eq:t0TI}
\end{equation}

It is instructive to consider Eq.~(\ref{eq:t0TI}) from a variety of 
perspectives.  A first question asks how the probability density for total 
duration changes on learning that the phenomenon is occurring:
\begin{equation}
p(T|I)=\int_{-\infty}^\infty dt_0\,p(t_0,T|I)={Tw(T)\over\overline T}\;.
\end{equation}
Notice that $p(T|I)$ is biased toward longer durations than the prior density 
$w(T)$.  This is because the phenomenon is very unlikely to be in progress 
at a random time selected from the long time interval $\Delta$, so finding 
it in progress prejudices you to think that it has a longer duration than 
your original expectations.

A useful, equivalent form for Eq.~(\ref{eq:t0TI}) comes from 
changing variables to present age and future duration.  The Jacobian of 
the transformation from $(t_0,T)$ to $(\tp,\tf)$ is $-1$, which 
implies that $dt_0\,dT=d\tp\,d\tf$.   Hence, the probability density for 
present age and future duration, given that the phenomenon is occurring, 
is
\begin{equation}
p(\tp,\tf|I)=p(t_0,T|I)
=\cases{w(\tp+\tf)/\overline T\;,&$\tp\ge0$ and $\tf\ge0$,\cr
	0\;,&otherwise.
       }
\label{eq:tptfI}
\end{equation}
Knowing the phenomenon is in progress is equivalent to saying that both the 
present age and future duration are nonnegative, so we can regard that 
condition as implicit and omit it from subsequent expressions.  The 
content of Eq.~(\ref{eq:tptfI}) is the following: if you know the
phenomenon is in progress, but don't know its present age, you treat 
uniformly the split of total duration into past and future; more precisely,
you assign the same probability, governed by $w(T)$, to all ways of splitting 
$T$ into past and future.  That's the temporal Copernican principle.  Indeed, 
{\it Eq.~(\ref{eq:tptfI}) is the mathematical embodiment of the temporal 
Copernican principle for phenomena known to be in progress.}  

Equation~(\ref{eq:tptfI}) has three immediate consequences that highlight 
the connection between the Copernican principle and Gott's rule.
We proceed by noting that once the phenomenon is known to be in progress, 
the total duration is the sum of the present age and the future duration,
i.e., $p(T|\tp,\tf,I)=\delta(T-\tp-\tf)$.  Another application of Bayes's
theorem then gives
\begin{equation}
p(\tp,\tf|T,I)={p(T|\tp,\tf,I)p(\tp,\tf|I)\over p(T|I)}
={1\over T}\delta(T-\tp-\tf)\;.
\end{equation}
This is a conditional version of Eq.~(\ref{eq:tptfI}), with the same content.

An obvious consequence is that if you know the phenomenon is in progress 
and also know its total duration, then you conclude that the present age 
is uniformly distributed between 0 and $T$:  
\begin{equation}
p(\tp|T,I)=\int_0^\infty d\tf\,p(\tp,\tf|T,I)=
\cases{1/T\;,&$\tp\le T$,\cr
       0\;,&$\tp>T$.
      }
\label{eq:uniformtpI}
\end{equation}
This is the precise statement of what Gott is trying to capture in his 
initial assumption~(\ref{eq:Copernicus}) about the present age.  The 
starting point~(\ref{eq:uniformtp}) of the flawed Bayesian analysis also 
asserts that $\tp$ is uniformly distributed between 0 and $T$, but it is 
different from Eq.~(\ref{eq:uniformtpI}) in a subtle, but crucial way: 
because $p(\tp|T,I)$ is conditioned on knowing the phenomenon is occurring, 
further statistical inference uses the conditional density 
$p(T|I)=Tw(T)/\overline T$, instead of the prior density $w(T)$; we 
find below [see Eq.~(\ref{eq:TtpI})] that this is how the present Bayesian 
analysis comes into agreement with the straightforward inference of the 
preceding section.

A second obvious, but important consequence of Eq.~(\ref{eq:tptfI}) is that
\begin{equation}
P\Biggl({1-b\over b}\tp<\tf<{1-a\over a}\tp\Biggm|T,I\Biggr)=
\int_0^T d\tp\int_{(b^{-1}-1)\tp}^{(a^{-1}-1)\tp}d\tf\,p(\tp,\tf|T,I)
=\int_{aT}^{bT}{d\tp\over T}=b-a\;.
\label{eq:tptfTba}
\end{equation}
Since the condition on future duration in the probability on the left is 
equivalent to $aT<\tp<bT$, this is just the statement that in dividing 
the total duration into past and future, possibilities satisfying the 
condition are a fraction $b-a$ of all the possibilities.  Furthermore, 
since the conditional probability~(\ref{eq:tptfTba}) is independent of $T$, 
the same result holds no matter what the prior density for $T$:
\begin{equation}
P\Biggl({1-b\over b}\tp<\tf<{1-a\over a}\tp\Biggm|I\Biggr)=
\int_0^\infty dT\,
P\Biggl({1-b\over b}\tp<\tf<{1-a\over a}\tp\Biggm|T,I\Biggr)p(T|I)=
b-a\;.
\label{eq:tptfba}
\end{equation}
Setting $b=1$ and $a=(1+Y)^{-1}$ in this result yields the third consequence 
of Eq.~(\ref{eq:tptfI}): 
\begin{equation}
P(\tf\ge Y\tp|I)=
{1\over1+Y}\;.
\label{eq:tfYtp}
\end{equation}
Equations~(\ref{eq:tptfba}) and~(\ref{eq:tfYtp}) are precise statements
of Gott's rule in the forms~(\ref{eq:Grule1}) and~(\ref{eq:Grule3}).  Indeed,
they look just like Gott's rule, with the crucial difference that they
are conditioned on knowing the phenomenon is in progress, without knowing
its present age.

We are now in the curious position of affirming that {\it for a phenomenon 
known to be in progress, but whose present age is unknown, the temporal 
Copernican principle leads to universal statistical predictions, which are 
described by Gott's rule}.  Indeed, all the manipulations in Gott's delta-$t$ 
argument are valid in this situation.  The down side for Gott is that this 
conclusion does not authorize his predictions: {\it in these circumstances, 
Gott's rule has no power to predict future durations from present ages, for 
the simple reason that the present age is unknown.}  

The results of the present Bayesian analysis make perfect sense in the
three examples introduced above.  Oddly enough, the deterministic example 
is the simplest: if you find the timer ticking, {\it but there is nothing 
to indicate how long it has been ticking}, it is reasonable to assign the 
same probability to all ways of dividing the 30-minute interval into past 
and future.  

The case of the women's longevity is a bit more complicated.  If you 
encounter an individual, {\it but are given no clue as to the individual's 
age}, a first cut might treat the past-future split uniformly.  For a 
person born in Britain, a more careful analysis would give greater weight 
to the future than to the past, because of the increase in life expectancy 
in this century.  I have already indicated that the case of an individual's 
longevity does not fit into the temporal Copernican principle for just 
this reason.  It should be emphasized that the problem is not the use of 
Bayesian analysis: the increase in life expectancy could be incorporated 
into a more complicated Bayesian analysis, which would automatically produce
a bias toward the future.

The case of the atom is particularly interesting because of Gott's 
exponential-decay derivation of Eq.~(\ref{eq:Grule3}).  If you find the atom 
in the excited state, {\it but you are not told when it was excited}, it is 
reasonable to assign the same probability to all ways of splitting a particular 
duration $T$ into past and future and to weight the result by an exponential 
$e^{-\lambda_0T}=e^{-\lambda_0(\tp+\tf)}$, which expresses the probability 
for duration $T$.  The final result, properly normalized, is the probability 
density~(\ref{eq:tptfI}) specialized to an exponential decay: 
\begin{equation}
p(\tp,\tf|I)=\lambda_0^2e^{-\lambda_0(\tp+\tf)}\;.
\label{eq:expdecay}
\end{equation}
This allows us to understand Gott's exponential-decay derivation of 
Eq.~(\ref{eq:Grule3}) [see Gott's Eq.~(6)]: he starts with 
Eq.~(\ref{eq:expdecay}) [see Gott's Eqs.~(3) and (4)], from which he 
immediately derives Eq.~(\ref{eq:tfYtp}), all without realizing that
Eq.~(\ref{eq:expdecay}) applies to an exponential decay whose present
age is unknown.  Having gone through a proper Bayesian analysis, we now 
understand that Eq.~(\ref{eq:tfYtp}) does not depend at all on assuming 
an exponential decay, but rather is a universal consequence of the temporal 
Copernican principle, valid no matter what the prior density $w(T)$,
provided the present age is unknown.

The next task is to find out what happens if you do discover the present age. 
When you determine the present age of the phenomenon, your Bayesian posterior 
for the total duration is given by
\begin{equation}
p(T|\tp,I)={p(\tp|T,I)p(T|I)\over p(\tp|I)}=
\cases{
	0\;,&$T<\tp$,\cr
	w(T)/Q(\tp)\;,&$T\ge\tp$,
       }
\label{eq:TtpI}
\end{equation}
where
\begin{equation}
p(\tp|I)=\int_0^\infty dT\, p(\tp|T,I)p(T|I)=Q(\tp)/\overline T\;.
\end{equation}
This posterior density is identical to the one that emerged from the 
straightforward Bayesian analysis that wholly ignored the Copernican 
principle.  This is as it should be, because in the language of this
section, the straightforward Bayesian inference corresponds to first 
learning the birth time $t_0$ and then discovering that the phenomenon 
has survived a time $t_p$, a situation that is equivalent to first 
learning that the phenomenon is in progress and then discovering its 
present age.  Once you are informed of the present age or, equivalently, 
of the birth time, you {\it are\/} at a special time, the time $\tp$ 
since the phenomenon began.  The temporal Copernican principle becomes 
irrelevant.  It just gets in the way of the obvious inference expressed 
in Eq.~(\ref{eq:rightpT}).  

At this point it is profitable to re-read Gott's account of his 1969 
encounter with the Berlin Wall.  If Gott had not known when the Wall 
was built, the logic of the first two paragraphs of his account would be 
impeccable.  Under those circumstances, it would be reasonable to assign 
probability 1/2 to the encounter's occurring during the middle two quarters 
of the Wall's total history.  Since he did know that the Wall was built in 
1961, however, his encounter did occur at a special time, the time eight 
years after the Wall's construction.  The predictions made in the third
paragraph of his account do not follow from the argument in the first
two paragraphs.  Indeed, his posterior expectations for the Wall's duration 
should have been a renormalized version of his prior expectations, whatever 
those were, with durations up to eight years excluded.

We can now give a succinct account of how Gott's delta-$t$ argument goes 
awry: the first two steps are wrong.  The step from Eq.~(\ref{eq:Copernicus}) 
to Gott's rule~(\ref{eq:Grule1}) is a non-Bayesian inference having no 
justification in probability theory; just as important, 
Eq.~(\ref{eq:Copernicus}) is itself an incorrect expression of the temporal 
Copernican principle, because it assumes that an observation at a random time 
will find the very unlikely result that the phenomenon is in progress.  In 
repairing these errors, we discovered that Gott's rule for relating future 
duration to present age is indeed a universal consequence of the temporal 
Copernican principle, but only in a situation---not knowing the present 
age---which leaves the rule shorn of predictive power.  Gott's predictions 
require knowing how long a phenomenon has lasted, but once you obtain this 
information, the temporal Copernican principle no longer has any impact, 
because you are at a special time within the lifetime of the phenomenon.

\bigbreak
 
\leftline{\bf Gott's rule as a predictor}
 
\nobreak 
\smallskip

All of Gott's predictions---from the future duration of the Berlin Wall to 
the longevity of the human species---are now detached from their original 
mooring in the temporal Copernican principle and left to float free of 
justification.  Yet a flawed analysis might lead to reasonable predictions.  
There might be some justification for Gott's predictions other than the 
Copernican principle.  Both the straightforward Bayesian analysis and the 
analysis based on the Copernican principle culminate in the same inference 
[Eqs.~(\ref{eq:rightpT}) and~(\ref{eq:TtpI})]: once you know the present 
age, your expectations about total duration are the same as your prior 
expectations, except that durations shorter than the present age are excluded.  
Thus all questions about the applicability of Gott's predictions reduce 
to determining what prior density underlies his predictions.

As noted above, Gott's rule follows from a posterior density
\begin{equation}
p(T|O)=\cases{0\;,&$T<\tp$,\cr
	      \tp/T^2\;,&$T\ge\tp$.
             }
\end{equation}
Within the correct Bayesian analysis, this posterior comes from an 
unnormalizable prior density 
\begin{equation}
w_g(T)={1\over T^2}\;.
\label{eq:wT2}
\end{equation}
This prior density, distinguished by a subscript $g$, corresponds to a survival 
probability $Q_g(T)=1/T$ and to a death rate $\lambda_g(T)=1/T$.  One way to 
characterize $w_g(T)$ is that the characteristic time associated with the 
death rate, $\lambda_g^{-1}(T)$, is always the same as the age $T$.

The prior density $w_g(T)$ is different from the Jeffreys prior that Gott 
\cite{Gott1994a,Gott1996a} identifies with his predictions, the reason being 
that Gott uses the flawed Bayesian analysis given above.  Yet within the
Bayesian analysis using the temporal Copernican principle, $w_g(T)$ has a 
scale-free status similar to that found by Jaynes \cite{Jaynes1968a} for
the Jeffreys prior.  Suppose that once you know the phenomenon is in progress, 
anything else you know, coming from the prior information about $T$, is 
unchanged by a simultaneous change in the scale of the past and the future.  
Under such a scale change, $\tp'=\alpha\tp$ and $\tf'=\alpha\tf$, the new 
and old probability densities are related by 
\begin{equation}
p'(\tp',\tf'|I)=p(\tp,\tf|I)\,d\tp\,d\tf/d\tp'\,d\tf'=p(\tp,\tf|I)/\alpha^2\;.
\end{equation}  
To say that all your information is unchanged by this scale change is to 
say that the old and new densities should have the same functional form, 
i.e., $p'(\tp',\tf')=p(\tp',\tf')$, which implies that 
$p(\alpha\tp,\alpha\tf|I)=p(\tp,\tf|I)/\alpha^2$.  Using Eq.~(\ref{eq:tptfI}) 
to write this in terms of the prior density, one finds that 
$w(\alpha T)=w(T)/\alpha^2$, which implies that the prior density has the 
form~(\ref{eq:wT2}). 

As discussed above, the Jeffreys prior applies when your prior information
about the duration, before any observation, is scale-invariant.  Once you
know the phenomenon is in progress, however, $w_g(T)$ captures the notion
of scale invariance, because it corresponds to invariance of $p(\tp,\tf|I)$
under simultaneous rescaling of the past and future.  In contrast, the 
Jeffreys prior corresponds to invariance of $p(\tp,\tf|I)$ under rescaling 
of $\tp$ or $\tf$, but not both simultaneously.

We have now uncovered the prior information that underlies the use of Gott's
rule as a predictor of future duration; namely, knowing that a phenomenon is 
in progress, you cannot identify any time scales associated with the phenomenon
either into the past or into the future.  One way of thinking about this is 
that for a phenomenon that has no time scales, discovering the present age
does {\it not\/} put you at a special time in the phenomenon's history, so 
some consequences of the temporal Copernican principle survive.  
Whether the scale-free prior information is appropriate must be judged case 
by case; it is not a universal rule.  The scale-free prior certainly does not 
apply to the three examples introduced in this article, each of which has 
an obvious time scale: for the atom, the scale is the decay time; for an 
individual, the scale is a typical human lifetime; for the deterministic 
phenomenon, the scale is the 30 minutes that the timer ticks.  Ignoring these 
time scales is the reason that Gott's rule leads to absurd predictions for 
these examples.
 
The examples Gott discusses at the beginning of his {\sl Nature\/} article
all have readily identifiable time scales that make application of Gott's 
rule problematic.  The survival of a human institution---a political 
institution such as the government of the former Soviet Union or a cultural 
institution such as a periodical like {\sl Nature}---is influenced 
by the 30-year time scale of a generation or by a typical human lifetime, 
since loyalty to and management of such institutions change on these time 
scales.  Physical manifestations of human institutions, such as the Berlin 
Wall or Stonehenge, are influenced by these same human time scales and, in
addition, by the time scale over which erosion leads to disintegration.

\bigbreak
 
\leftline{\bf The success of Gott's rule}
 
\nobreak 
\smallskip

Even though there is little reason to adopt Gott's rule, he portrays
his predictions as successful \cite{Ferris1999a,Gott1993a,Gott1994a,Gott1996a}.  
Consider, for example, his 95\%-confidence prediction that {\sl Nature}, 
given its 123-year history of publication in 1993, would continue to publish 
for a period between 3.15 years and 4,800 years.  Gott would consider this 
prediction successful because {\sl Nature\/} has already surpassed the lower 
bound and is very unlikely to exceed the upper bound.  Yet there's the hitch: 
the upper bound is far too large; without doing any analysis, anyone could 
have written down a similar very large 95\% confidence interval and achieved 
the same ``success.''  To assess Gott's rule, one should direct attention not 
at the the 95\% confidence predictions, but at the high probabilities the 
rule assigns to very long future durations.  Gott's rule in the 
form~(\ref{eq:Grule3}) predicts that with probability 1/2, {\sl Nature\/} 
will continue to publish for more than 123 years after 1993, with probability 
1/5 for more than 492 years, with probability 1/10 for more than 1,107 years, 
and with probability 1/20 for more than 2,337 years.  These probabilities posit 
a great deal of faith in the durability of human institutions.

To make this point more quantitatively, it is useful to consider a particular 
form of the probability that future duration exceeds some multiple of present 
age: 
\begin{eqnarray}
P(\tf\ge Y\tp|O)&=& 
P\Bigl(T\ge(1+Y)\tp\bigm|O\Bigr)\nonumber\\
&=&\int_{(1+Y)\tp}^\infty dT\,p(T|O)
={Q[(1+Y)\tp]\over Q(\tp)}
=\exp\!\left(-\int_{\tp}^{(1+Y)\tp}dT\,\lambda(T)\right)\;.
\label{eq:survival}
\end{eqnarray}
This form makes clear that $P(\tf\ge Y\tp|O)$ depends only on the death rate 
during the interval between the present age and the lower bound for longevity.  
For a death rate $\lambda_g(T)=1/T$, one gets Gott's rule.

Now let's apply this to the example of a periodical like {\sl Nature}.  At 
start-up a new publication confronts a variety of short-term, rapid-death 
scenarios.  Should it survive these initial hazards and become established
like {\sl Nature}, the next time scale it faces might be roughly a human 
lifetime.  If this time scale is modeled by a constant death rate 
$\tau^{-1}=(60\,{\rm yr})^{-1}$, then one finds from Eq.~(\ref{eq:survival}) 
that $P(\tf\ge Y\tp|O)=e^{-Y\tp/\tau}$.  For {\sl Nature\/}, this gives 
predictions quite different from Gott's: for example, a probability $0.129$ 
to continue publishing for more than 123 years beyond 1993 and a probability  
$2.75\times10^{-4}$ to continue publishing for more than 492 years.

Should these predictions seem unduly pessimistic, it is because the constant
decay rate does not recognize a long publication record as providing evidence 
for continued success.  A prejudice that success begets success can be 
incorporated, without discarding the time scale, by choosing, for example,
\begin{equation}
\lambda(T)={1\over T}\left({T\over\tau}\right)^\beta\;,
\end{equation}
where $\beta\ge0$.  For $\beta=1$, this gives a constant death rate 
$\tau^{-1}$, and for $\beta=0$, it gives Gott's rule.  For intermediate 
values, it gives a death rate that decreases with age, but with the 
time scale $\tau$ still having an effect.  The resulting 
probability~(\ref{eq:survival}) is 
\begin{equation}
P(\tf\ge Y\tp|O)=
\exp\!\left[
-\left({\tp\over\tau}\right)^\beta{(1+Y)^\beta-1\over\beta}
\right]\;,
\end{equation}
For {\sl Nature\/} this gives, assuming $\beta=1/2$, a probability $0.305$
to continue publishing for more than 123 years beyond 1993, a probability
$2.90\times10^{-2}$ for more than 492 years, and a probability 
$2.05\times10^{-3}$ for more than 1,107 years.  The point here is not the 
particular values nor even the death-rate model, but rather that there is 
one or more time scales, which can and should be incorporated in the prior 
distribution. 

Gott stresses the success of his predictions \cite{Ferris1999a,Gott1996a} 
for the 44 Broadway and off-Broadway plays listed in {\sl The New Yorker\/} 
on 27~May~1993, the day his original {\sl Nature\/} article was published. 
For example, Gott's 95\%-confidence rule predicted that {\sl Cats}, having 
played for 3,885 days, would continue to play for a period between 100 days 
and 415 years.  Gott regards this prediction as a success because the 
production continues today, thereby surpassing the lower bound, and is 
unlikely to exceed the upper bound \cite{Ferris1999a}.  Yet since {\sl Cats\/} 
had run 6,263 days through 30~November~1999, when I determined that it was 
still running, the same rule predicts that with probability 1/5, it will 
continue to run for at least another 68.6 years, with probability 1/10 for 
at least another 154 years, and with probability 1/20 for another 326 years.  
Such predictions ignore obvious time scales.  A new production faces a variety 
of short- to medium-term scales, including the time to the first reviews, 
the time over which a producer is willing to back a losing production, the 
annual cycle of openings and closings, and the time over which a star performer 
tires of a particular part and moves on to other challenges.  An established 
production like {\sl Cats}, having survived these initial hurdles, must deal 
with the decade- to generation-long scale over which taste and fashion change 
substantially and the production experiences a nearly 100\% turnover of
personnel.  Including this long-term scale would temper Gott's predictions 
for extraordinarily long runs.

The problems with Gott's long-term predictions show up more dramatically 
in phenomena, such as the longevity of an individual, where an initial 
period of low death rate is followed by relatively rapid extinction.  We
don't need a detailed model to tell us whether we should believe Gott's
prediction, based on his age of 46.3 years on 27~May~1993, that he has a
1/3 chance to survive to more than 139 years old.

There are two reasons, in my view, why Gott is able to get away with making
his scale-free predictions for the survival of governments and plays and
periodicals.  First, statistical models for the longevity of these phenomena 
are not well developed, so Gott is protected from the absurdities that arise 
immediately in the three examples used in this article.  Although there are 
readily identifiable time scales associated with the phenomena Gott considers, 
how to incorporate them into prior probabilities for duration has not been 
much investigated.  There's a good reason for this: to assess the viability 
of an established government or play or periodical, readily available current 
data about the particular phenomenon in question---data such as the popularity 
of the government, the balance sheet of the play or periodical, trends in 
attendance at the play or the number of subscribers of the periodical---are 
far more cogent than prior information about longevity together with the
present age.  Second, the intervals that Gott finds for survival times are 
so wide that he is likely to be right, till he is forced to place bets based 
on the high probabilities he assigns to long survival times.   A negative 
feature of such bets, however, is that the bettors might not survive till 
the bets are settled.  Even for the case of human longevity, where one 
could easily formulate bets that Gott would almost certainly lose, the time 
scales are long enough that one might not get much personal satisfaction 
from winning.

A way to overcome this difficulty is to bet on the survival of creatures
with a shorter lifetime than humans, but for which data on present age
and future survival are readily obtainable.  For this purpose, I sent an
e-mail on 21~October~1999 and again on 2~December~1999, to my department's 
most comprehensive e-mail alias, which includes faculty, staff, and graduate 
students, requesting information on pet dogs.  The responses were compiled 
and checked for accuracy on 6~December; a notarized list of the 24 dogs, 
including each dog's name, date of birth, and breed, and the caretaker's 
name, was deposited in my departmental personnel file on 21~December~1999.  
Gott's rule predicts that each dog will survive to twice its present age with 
probability 1/2.  For each of the 6 dogs above 10 years old on the list, 
I am offering to bet Gott \$1,000\,US, at odds of 2:1 in his favor, that the 
dog will not survive to twice its age on 3~December~1999.  The reason for 
weighting the odds in Gott's favor is to test his belief in his own 
predictions: given the odds, his rule says that his expected gain, at 
\$1,000 per bet, is \$6,000; moreover, the probability that he will be a 
net loser (by losing five or more of the six bets) is 7/64=0.109.

\bigbreak
 
\leftline{\bf Discussion}
 
\nobreak
\smallskip

The stated objective of this article is to determine what prior information
underlies Gott's rule.  Gott proposed his rule as a predictor of future 
duration based on knowing the present age and nothing else.  What we have 
discovered is that the actual prior information underlying Gott's rule is 
both less and more than he thought.  On the one hand, Gott's rule is a 
consequence of the temporal Copernican principle for a phenomenon whose 
age is unknown, but this universal form of Gott's rule has no predictive 
power for future durations.  On the other hand, Gott's rule as a predictor 
of future durations is a consequence of discovering the present age of a 
phenomenon that has no identifiable time scales in the past or future.

What about the focus of Gott's {\sl Nature\/} article, the longevity of the
human species?  A species's survival depends on its ability to adapt to 
short- and long-term environmental changes produced by other species in 
its ecosystem and by climatological and geological processes.  The adaptations
are made possible by existing genetic variability in the gene pool and by 
random mutation.  How {\it homo sapiens\/} fits into this picture is a 
complicated question, certainly not amenable to a universal statistical rule.  
As Ferris \cite{Ferris1999a} puts it, ``\,\dots\,in my experience most people 
either think we're going to hell in a handbasket or assume that we're going 
to be around for a very long time.''  Both views are a reflection of 
advancing technology.  The first comes from alarm at technology's increasing 
impact---changes might be so rapid that we (and certainly other) species 
could not adapt.  The second comes from a belief that technology can save 
us---by controlling the environment or by making possible remarkable 
adaptations such as escaping our earthly environment or changing our 
genetic constitution.  

Gott dismisses all such thinking as the illusions of those who don't appreciate 
the power of the Copernican principle.  He contends that everything relevant 
to assessing our future prospects is contained in the statement that we are 
not at a special time.  This article shows that the Copernican principle
is irrelevant to considerations of the longevity of our species.  Perhaps we 
are still subject to the factors that determine the survival of other species.  
More likely, our survival---and the survival of many other species along
with us---depends on what we do now and in the future.  We better think hard 
about it.

\vspace{18pt}

\begin{quotation}
\small
{\sl Carlton M. Caves\/} has been Professor of Physics and Astronomy at the 
University of New Mexico since 1992.  He received a Ph.D.~in Physics from 
the California Institute of Technology in 1979, working mainly on the theory 
of gravitational-wave detection.  His subsequent career has proceeded from 
relativity theory through quantum optics to quantum information science, the 
unifying strand being an interest in how quantum mechanics impacts what we 
can measure and what we can know.  His interest in probability theory springs 
from the conviction that to have any hope of understanding the weird features 
of quantum mechanics, one must first have a clear idea of what probability 
means.  He maintains that the only consistent way of thinking about 
probabilities is the Bayesian interpretation, which holds that probabilities, 
far from being physical properties, are a measure of credible belief based 
on what one knows.  His main accomplishment is to have had a series of 
excellent Ph.D.~students and postdocs, who continue to teach him most of 
what he knows.
\end{quotation}

\end{document}